\documentclass[prl,twocolumn,amsmath,amssymb,showpacs,superscriptaddress]{revtex4-1}

\usepackage{amsfonts}
\usepackage{amssymb}
\usepackage{amsmath}
\usepackage{graphicx}
\usepackage{soul}
\usepackage[table,xcdraw]{xcolor}
\usepackage{dcolumn}
\usepackage{epstopdf}
\usepackage{bm}
\usepackage[T1]{fontenc}
\usepackage{hyperref}
\usepackage{wasysym}
\definecolor{Gray}{gray}{0.89}

\begin{document}

\title{Composite fermions in graphene fractional quantum Hall state at half filling: evidence for Dirac composite fermions}

\author{Antti Laitinen}
\affiliation{Low Temperature Laboratory, Department of Applied Physics, Aalto University, Espoo, Finland}
\author{Manohar Kumar}
\affiliation{Low Temperature Laboratory, Department of Applied Physics, Aalto University, Espoo, Finland}
\author{Pertti J. Hakonen}
\affiliation{Low Temperature Laboratory, Department of Applied Physics, Aalto University, Espoo, Finland}

\begin{abstract}

Composite fermions in fractional quantum Hall (FQH) systems are believed to form a Fermi sea of weakly interacting particles at half filling $\nu=1/2$. Recently, it was proposed (D. T. Son, Phys. Rev. X \textbf{5}, 031027 (2015)) that these composite fermions are Dirac particles. In our work, we demonstrate experimentally that composite fermions found in monolayer graphene are Dirac particles at half filling. Our experiments have addressed FQH states in high-mobility, suspended graphene Corbino disks in the vicinity of $\nu=1/2$. We find strong temperature dependence of conductivity $\sigma$ away from half filling, which is consistent with the expected electron-electron interaction induced gaps in the FQH state. At half filling, however, the temperature dependence of conductivity $\sigma(T)$ becomes quite weak as expected for a Fermi sea of composite fermions and we find only logarithmic dependence of $\sigma$ on $T$. The sign of this quantum correction coincides with weak antilocalization of composite fermions, which reveals the relativistic Dirac nature of  composite fermions in graphene.

\end{abstract}


\maketitle

The fractional quantum Hall (FQH) state  is a many body phenomenon where fractionally charged elementary excitations
lead to quantization of the Hall conductance at fractional filling factor $\nu = hn/(eB)$ at carrier density $n$ and magnetic field $B$  \cite{Tsui1982}. The generation of these incompressible liquid states requires a large Coulomb interaction energy compared with the disorder potential, putting strict  requirements on temperature,  the quality of the two-dimensional electron gas (2-DEG) and the strength of the magnetic field. Owing to reduced screening in atomically thin graphene, the electrons in graphene interact with  larger Coulomb interaction energy than electrons in semiconductor heterostructures, providing an extraordinary  setting for studies of FQH states and their description in terms of composite fermions \cite{Jain1989,Jain2007,Jain2015}.

The composite fermion theory \cite{Jain1989} and Composite Fermion Chern-Simon (CFCS) theory have been very successful in outlining a unified picture of fractional quantum Hall effect.
Lopez and Fradkin \cite{Lopez1991} showed that the problem of interacting electrons moving in 2D in the presence of an external magnetic field is equivalent to a fermion system, described by a Chern-Simon gauge field, where electrons are bound to even number of vortex  lines. Fluctuations in the gauge field were soon realized to have strong influence on the quantum correction of the composite fermion conductivity \cite{kalmeyer1992}. Subsequently, a Fermi liquid type of theory was proposed for half-filled Landau level  \cite{Halperin1993} where various observables in the low-temperature limit are described  in terms
of Fermi liquid parameters \cite{Simon1998},  involving most notably the effective mass $m^*$ for composite fermions, which is expected to have a strong enhancement near half filling.

There is extensive experimental evidence in favor of weakly interacting Fermi sea of composite fermions effectively in a zero magnetic field at $\nu=1/2$. Transport anomalies in the lowest Landau level of two-dimensional electrons at half filling were observed by Jiang et al. \cite{Jiang1989}. Distinct features related with compressibility  in surface-acoustic-wave propagation on high-quality AlGaAs/GaAs heterostructures were observed by Willett et al. \cite{Willett1990,Willett1993,Willett1998}. Furthermore,  resonances at fields where the classical cyclotron orbit becomes commensurate with a superlattice have been found \cite{Kang1993,Shayegan2016}. Strong enhancement of composite fermion mass near half filling $\nu =1/2$ has been observed in experiments on similar 2D electron gas heterostructures
\cite{Manoharan1994,Du1994a}.  Logarithmic temperature dependence of conductivity at half-integer filling factors has been observed, which has been interpreted to point towards residual interactions between composite fermions \cite{Kang1995,Rokhinson1995,Rokhinson1997}.

A particle-hole symmetric model for the Fermi liquid ground state of a half-filled Landau level has recently been reinvestigated \cite{Son2015}, with the conclusion  that composite fermions in a 2-DEG should, in fact, be Dirac particles. This conclusion has separately been verified by extensive numerical renormalization group analysis by Geraedts et al. \cite{Geraedts2016}. As Dirac particles, composite fermions may pick up a Berry phase when traversing around a closed loop, which has significant implications on their backscattering dynamics. Reduced backscattering will lead to weak antilocalization (WAL) which has been observed for Dirac particles in graphene at small magnetic fields \cite{Tikhonenko2009}. Similar behavior can be anticipated also for composite fermions in graphene acting as Dirac particles.

 In this work, we investigate quantum corrections to conductivity for a half-filled quantum Hall state in monolayer graphene sample which is formed of a suspended Corbino disk with high mobility. We find that composite fermions in graphene are Dirac particles as evidenced by the observed weak (anti) localization behavior at half filling. Our results display typical logarithmic temperature corrections for two-dimensional  weak localization (WL), but we find an opposite sign for these quantum corrections with respect to previous observations in 2D electron gas experiments \cite{Rokhinson1995,Rokhinson1997}.

Suspended graphene Corbino disks provide a clear-cut setting to probe  composite fermion physics in graphene at high fields.  Recently, Corbino geometry was  used in graphene experiments  which  focused on studies of magneto-conductance  in the quantum Hall regime~\cite{Zhao2012,Peters2014}. These previous measurements on graphene Corbino disks have failed to show any fractional states, perhaps due to strong charge inhomogeneity induced by the substrate. Our experiments at mK temperatures, however, display a multitude of FQH states (up to 11 states) in the lowest Landau level which are clearly visible in magneto- and transconductance measurements \cite{Kumar2016}. In our devices, conductance is governed by bulk properties instead of the  chiral edge states of regular quantum Hall bars. Consequently, our current-annealed, suspended graphene Corbino disks form an excellent platform to investigate the dynamics of the free Fermi sea of composite fermions at half filling.

\begin{figure}[tbp]
\includegraphics[width=0.43\textwidth]{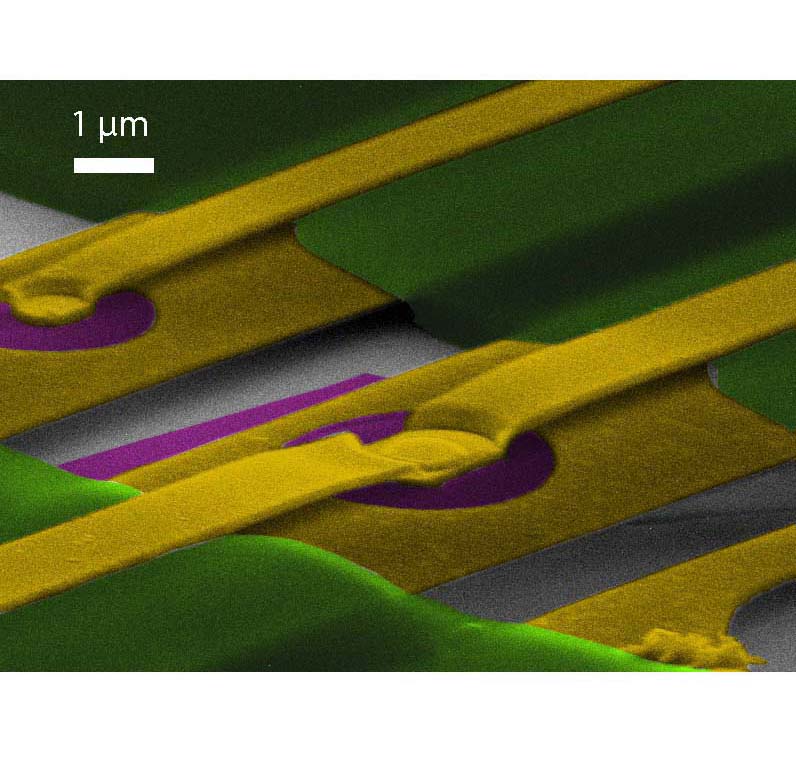}
\includegraphics[width=0.5\textwidth]{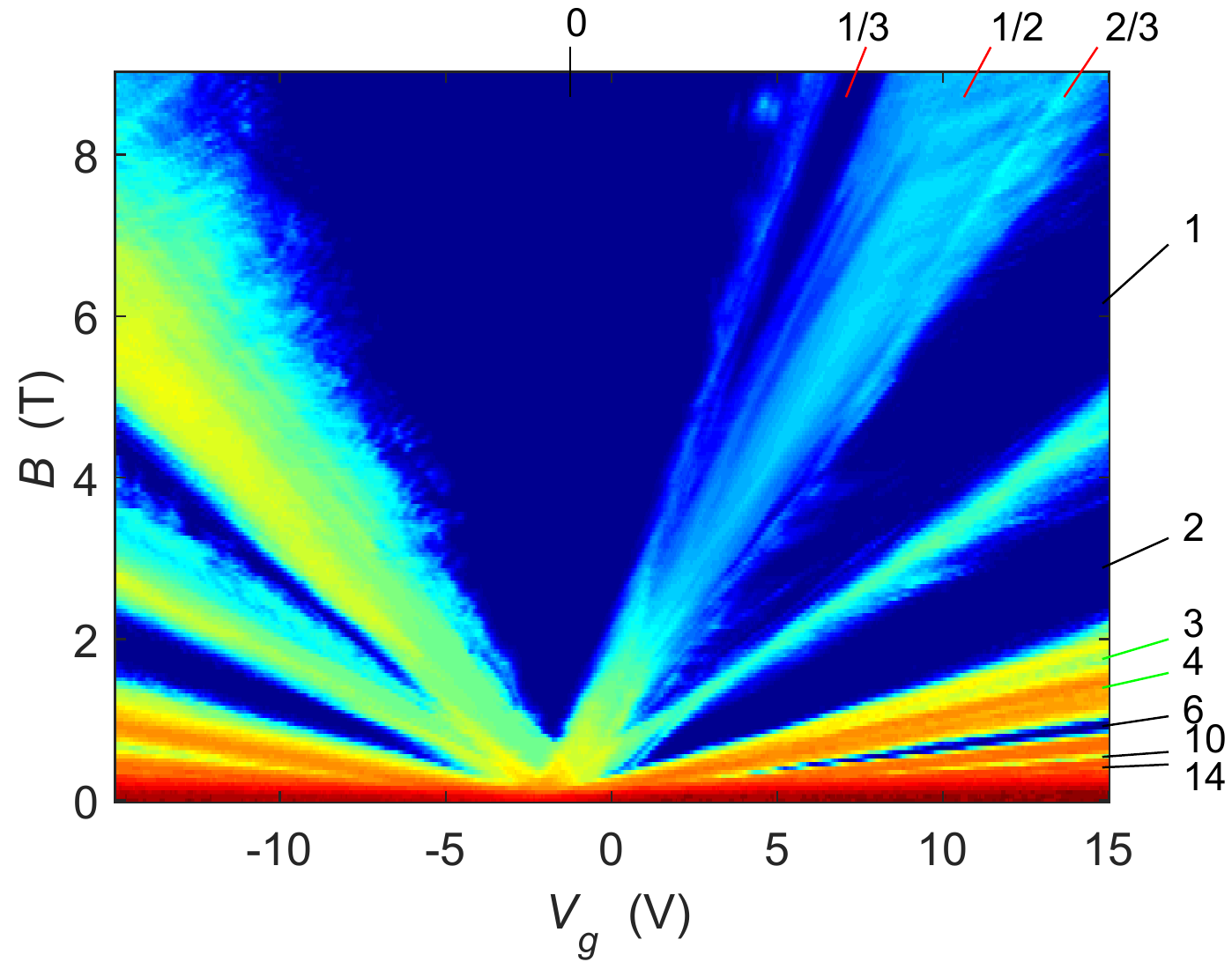}
\caption{ (a) Scanning electron micrograph of a 3.8-$\mu$m-diameter graphene Corbino sample (EV2) with 1500-nm diameter circular middle contact (in the center of the image); purple denotes the suspended graphene while gold appears as yellow. The second investigated sample EV3 is seen on the left. The white scale bar corresponds to 1 $\mu$m.  (b) The magneto-conductance of sample EV2 measured using AC peak-to-peak  excitation current  $I_{p-p} = 0.1$ nA at $20$ mK, displayed as a Landau fan diagram on the $B$ vs. gate voltage $V_g$ plane. The region of half filling is given by the slope 0.64 T/V starting from zero-field Dirac point at $V_g = -2.3$ V.
}

\label{sample}
\end{figure}

Interaction and weak localization corrections to conductivity in graphene are influenced by the pseudospin introduced by the two atom basis of the hexagonal lattice. The pseudospin leads to an additional Berry phase of $\pi$  in the backscattering wave interference, resulting in reduced back scattering or, in other words, antilocalization \cite{Suzuura2002a,Khveshchenko2006,Mccann2006}.  Real time reversal symmetry of graphene requires intervalley scattering which makes its weak localization phenomena intriguing. In fact, low field magnetoresistance in graphene may be positive or negative, depending on the strength of intervalley and dephasing scattering, as well as the scattering strength for TRS (time reversal symmetry) breaking  in a single valley \cite{Mccann2006,Tikhonenko2009,Couto2014}.

A charge carrier is combined with two vortices in the composite fermion picture. The underlying valley structure will be reflected on the graphene composite fermions. Hence, we expect similar scattering processes among graphene composite fermions as for graphene Dirac particles at small fields. Consequently, magnetoresistance at $\nu=1/2$ can be either positive or negative depending on  the rates of intervalley, dephasing and single-valley TRS breaking scattering for composite fermions denoted as $\tau_i^{-1}$, $\tau_{\varphi}^{-1}$, and $\tau_{*}^{-1}$, respectively.
Even though fluctuations due to Chern-Simon fields will enhance the dephasing rate of composite carriers, we expect the tendency towards weak antilocalization to remain, which reflects the presence of a Berry phase due to the Dirac nature of these composite carriers. Our results do confirm this expectation and display weak antilocalization behavior for graphene composite fermions.

\section{Samples and basic experimental results}

The samples considered here are denoted by EV2 ($r_0$ = 1900 nm, $r_i$ = 750 nm) and EV3 ($r_0$ = 1600 nm, $r_i$ = 400 nm), where $r_o$ and $r_i$ denote outer and inner radii of the Corbino disks. Our sample fabrication is explained in detail in Ref. \onlinecite{Kumar2016}.  We exfoliated graphene using a heat-assisted exfoliation technique to maximize the size of the exfoliated flakes~\cite{Huang2015}.  The contacts were deposited  by developing further  the ideas presented in Ref. \onlinecite{Tombros2011}. Strongly doped silicon Si++ substrate with 285 nm of thermally grown SiO$_2$  was used as a global back gate. Annealing of samples on LOR was typically performed at a bias voltage of 1.6$\pm$0.1 V which is comparable with our HF etched, rectangular two-lead samples \cite{Laitinen2014}. Larger Corbino disks can require higher annealing voltages, as was the case with sample EV3 where $\simeq2$ V was used.

Our suspended graphene  structure is illustrated in Fig. \ref{sample}a: the Corbino disk is supported only by the inner and outer  Au/Cr leads.  The conductivity was calculated from the measured conductance $G$ using $\sigma = \frac{G}{2\pi}\ln(r_o/r_i)$. The field-effect mobility was determined using equation  $\mu_f = \frac{\sigma-\sigma_0}{ne}$ after subtracting out the contact resistance from $\sigma$ and the measured minimum conductivity $\sigma_0$ at the Dirac point ($ V_g=V_g^d \sim -2$ V). Both samples had $\mu_f \simeq 2 \times 10^5$ cm$^2$/Vs. The residual charge density $n_{0} \simeq  6\times 10^9$ $1/\textrm{cm}^2$ was identified by looking for a cross-over between constant and power law behavior in $\log G$ \emph{vs.} $\log n$ traces.  Contact doping \cite{Laitinen2016} is estimated to correspond to a charge density of $n_c \sim 5 \times 10^{11}$ $1/{\textrm{cm}}^2  $ under the contact metal. Details concerning determination of basic parameters are given in Ref. \onlinecite{Kumar2016}.

In zero magnetic field, the  conductance of graphene in Corbino geometry at $V_g^d$ equals to $G=\frac{8e^2}{h} \log(\frac{r_o}{r_i})^{-1}$ according to conformal mapping theory \cite{Rycerz2009}. After subtraction of the contact resistance, our measured conductivity  is in line with the above  theoretical value due to evanescent modes. Also the measured gate voltage dependence of $G(V_g)$ in the unipolar regime was found to agree  with theoretical formula, which at the same time gave an estimate $R_c$ = 410 $\Omega$ for the contact resistance. The high quality of our samples is also manifested in the observability of broken symmetry states ($\nu =0, 3, 4$) down to 0.6 Tesla.

Our measurements down to 20 mK were performed on a BlueFors LD-400 dilution refrigerator. The measurement lines were twisted pair phosphor-bronze wires supplemented by three stage $RC$ filters with a nominal cut-off given by $R=100$ Ohms and $C=30$ nF. However, due high impedance of the quantum Hall samples the actual cutoff is determined by the sample resistance.  For magneto-conductance measurement in Fig. \ref{sample}b, we used an AC peak-to-peak current excitation of $0.1$ nA at $f=3.333$ Hz. The other magnetoconductance measurements were conducted using DC with sufficient pause times, and thus no signal suppression by cut-off is present. Due to symmetry reasons of the Corbino geometry, the azimuthal electric field is zero. Consequently, our experiment measures directly the longitudinal conductivity $\sigma_{xx}$ of our sample.

\section{Magnetoconductance around 1/2 filling}

\begin{figure}[tbp]
\includegraphics[width=0.49\textwidth]{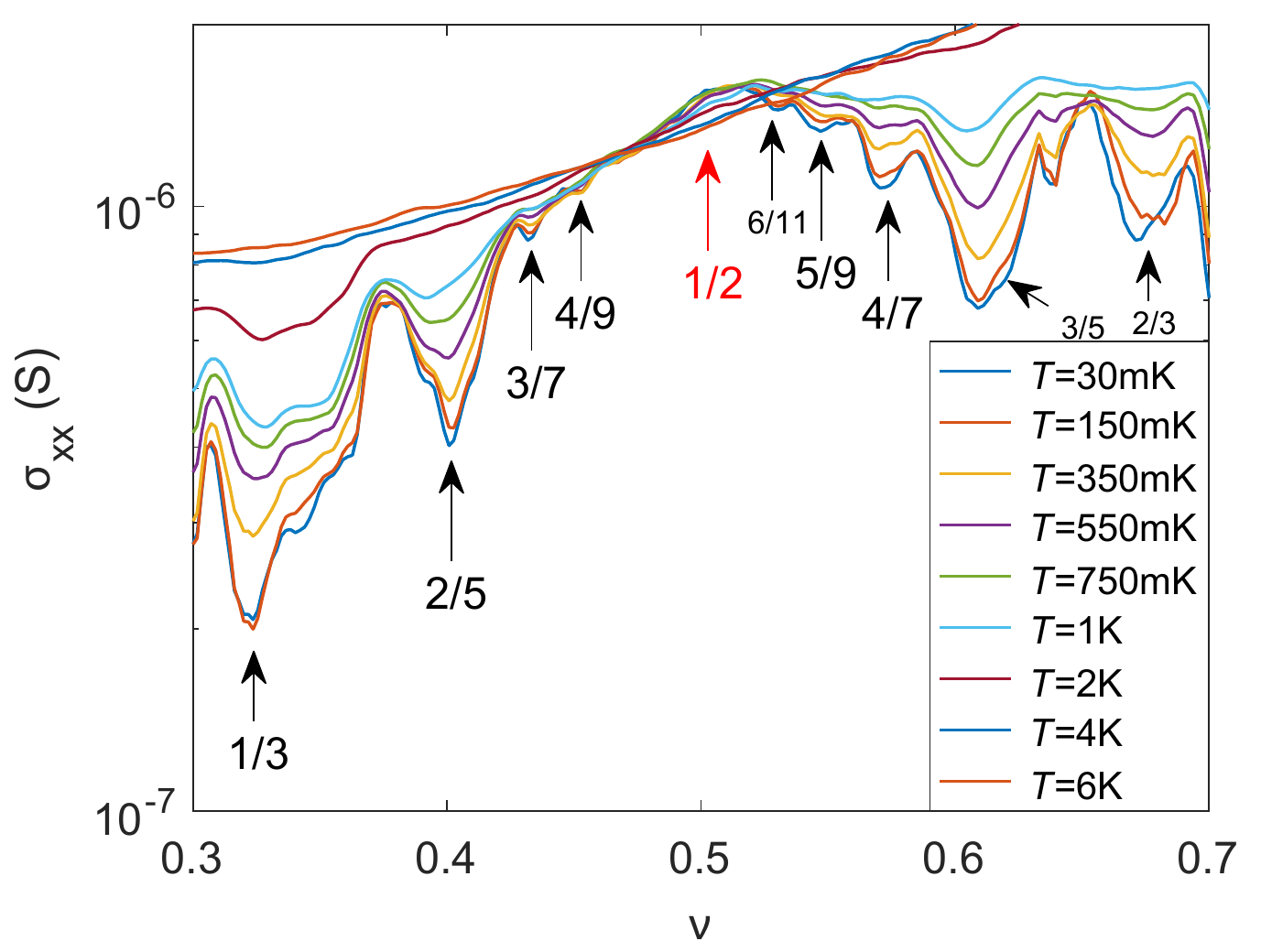}
\includegraphics[width=0.49\textwidth]{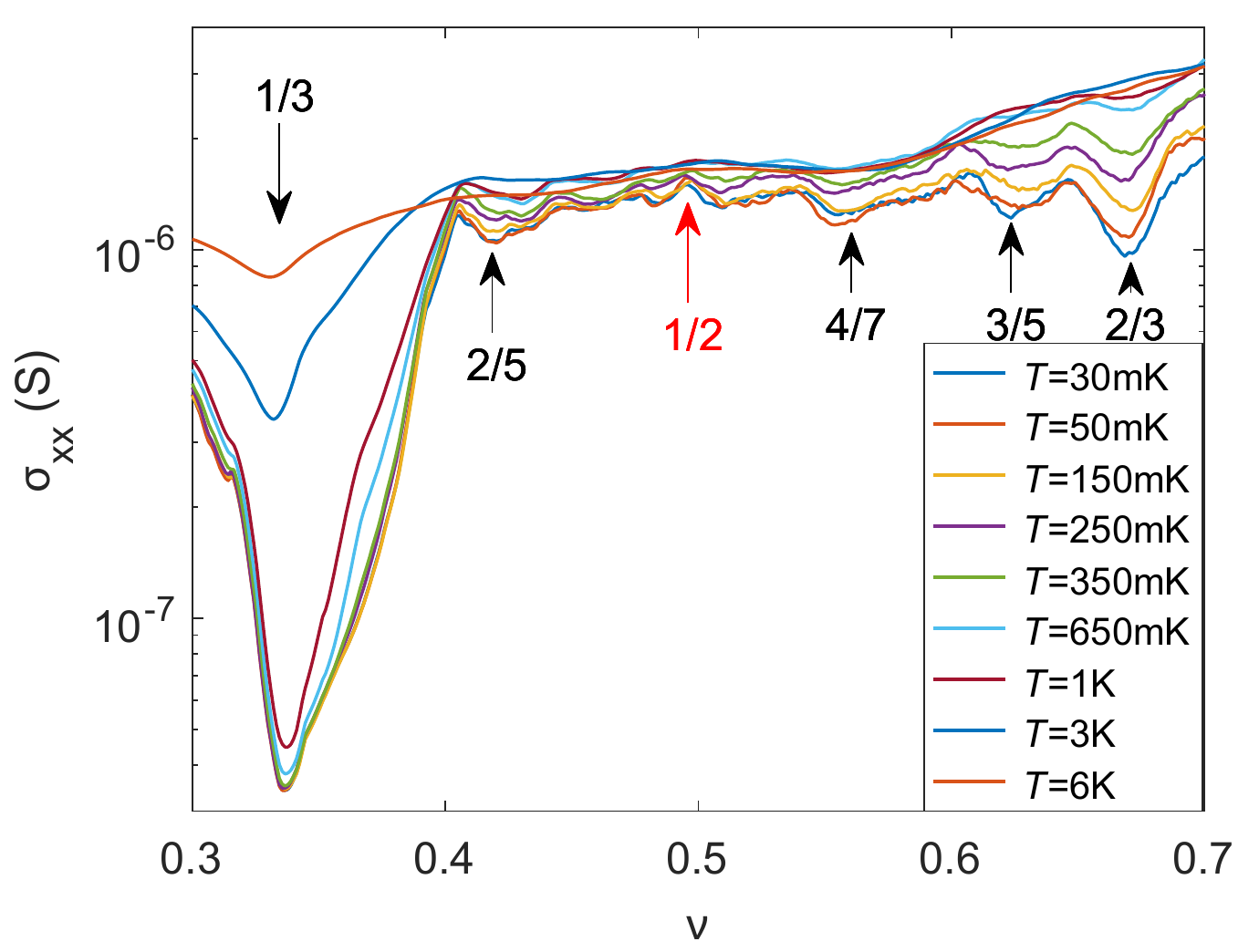}

\caption{a) Temperature dependence of conductivity for Corbino sample EV2 measured over charge density $n$ = 3.8 - 7.5 $\times 10^{10}$ $\frac{1}{\textrm{cm}^2}$ ($V_g$ up to 6.9 V) at $B = 5$ T. FQH-states are marked using black arrows. b) Data at $B$ = 9 T  for the same sample EV2. The charge carrier range here covers $n$ = 6.5 - 15.2 $\times 10^{10}$ $\frac{1}{\textrm{cm}^2}$ ($V_g$ up to 13.9 V)}
\label{Tdep}
\end{figure}

Our data on electronic conductivity $\sigma_{xx}$ vs. gate-swept filling factor $\nu = hn/(eB)$   is displayed in Fig. \ref{Tdep}  for several temperatures in the range $T = 0.03 - 6$ K measured at $B=5$ T and 9 T. Half filling regime is characterized by the weakest temperature dependence of $\sigma_{xx}$, which is in accordance with the absence of any energy gap in this regime. In both frames, only linear background variation of $\sigma_{xx}(\nu)_{|_{6\textrm{K}}}$ around $\nu=1/2 \equiv hn/eB_{1/2}$ is observed  at $T=6$ K, which indicates washing out of quantum corrections of conductance due to strong dephasing.

At 5 T (see Fig. \ref{Tdep} a), we find a change in effective magnetoconductance defined as $\delta \sigma_{xx}^{eff} = \sigma_{xx}(\nu)_{|_T} - \sigma_{xx}(\nu)_{|_{6\textrm{K}}}$, which varies between  positive and negative values with $n$ around $\nu =1/2$ at low temperatures. We interpret this as varying sign of $\delta \sigma_{xx}^{eff} $ as coming from a competition between WL/WAL behavior and the appearance of gaps of the fractional states at $\nu \neq 1/2$. This complicated localization behavior is assigned to fluctuations in Chern-Simon fields and the particular band structure of graphene influencing its weak localization properties \cite{Mccann2006}.
By enhancing the applied magnetic field, the effective magnetoconductance $\delta \sigma_{xx}^{eff}$  in sample EV2 acquires a negative sign around $\nu=1/2$ at fields of $7.5-9$  T, as illustrated by the data in Fig. \ref{Tdep} b (in sample EV3 $\delta \sigma_{xx}^{eff} < 0$ at $B=5-9$ T). The half-filling data at 9 T displays $d\sigma/dT > 0$ and $d\sigma/dB < 0$, which form the fundamental findings before conversion to the composite fermion description.

The temperature dependence in Fig. \ref{Tdep} can also be employed to determine energy gaps $E_g/k_B$ around $\nu=1/2$, for which we find at 5T $(B - 2hn/e) \times 3$ K/T $-\Gamma_0$  and $(B - 2hn/e) \times 6$ K/T $-\Gamma_0$  with $\Gamma_0 \sim 1-2 $ K  for particles and holes, respectively. Using the slopes $dEg/d\nu$, we obtain for the effective masses $0.4m_e$ and $0.8m_e$, respectively, where $m_e$ denotes the electron mass.

The effective field seen by the composite fermions at half filling is $B^{*} = B - B_{1/2}=0$.  To compare the magnetoconductivity in Fig. \ref{Tdep}  with phenomena at small fields, we note that there may be either positive or negative magnetoresistance in graphene \cite{Mccann2006}. The sign of
magnetoresistance has been demonstrated to depend fundamentally on parameters $\tau_{\varphi}/\tau_i$ and $\tau_{\varphi}/\tau_*$; when both of these values are $\lesssim 1$, weak antilocalization is preferred and $d\sigma/dT < 0$ \cite{Tikhonenko2009}. In Ref. \onlinecite{Tikhonenko2009}, small ratios were achieved at elevated temperatures $T >> 4.2$ K. In our suspended sample, we were able to observe WAL even  at 20 mK, and the range of  weak antilocalization magnetoresistance was found to  be within $\delta B \simeq 20$ mT. This makes field-swept investigations of weak localization at $|B^*| < \delta B$ challenging in our graphene samples, primarily since we have a charge inhomogeneity corresponding to variation $\delta \nu / \nu \gtrsim \delta B /B$. Nevertheless, we  see  clear quantum corrections in conductivity near half filling and these corrections depend logarithmically on  temperature.

\section{Quantum corrections}

\begin{figure}[tbp]
\includegraphics[width=0.47\textwidth]{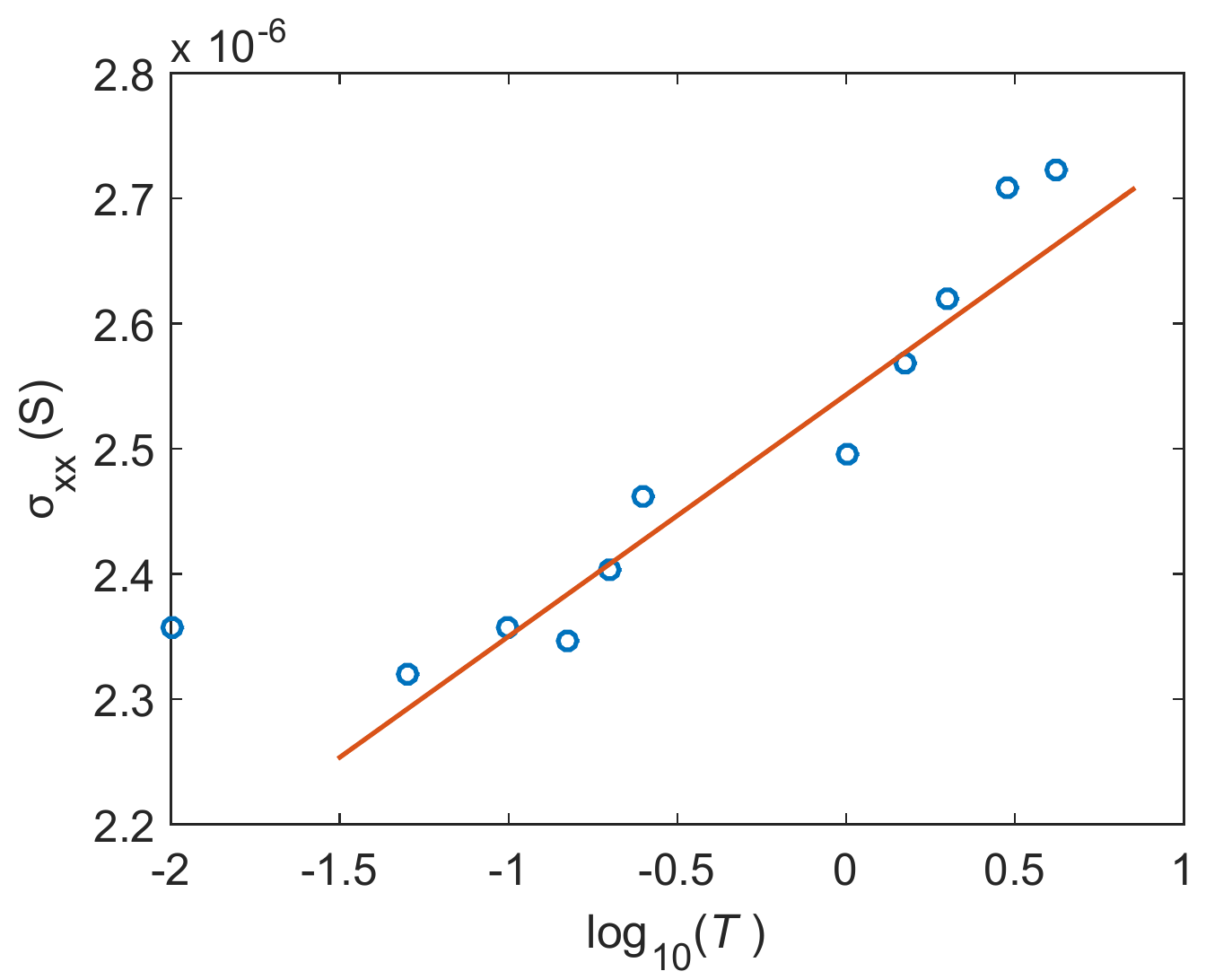}
\includegraphics[width=0.47\textwidth]{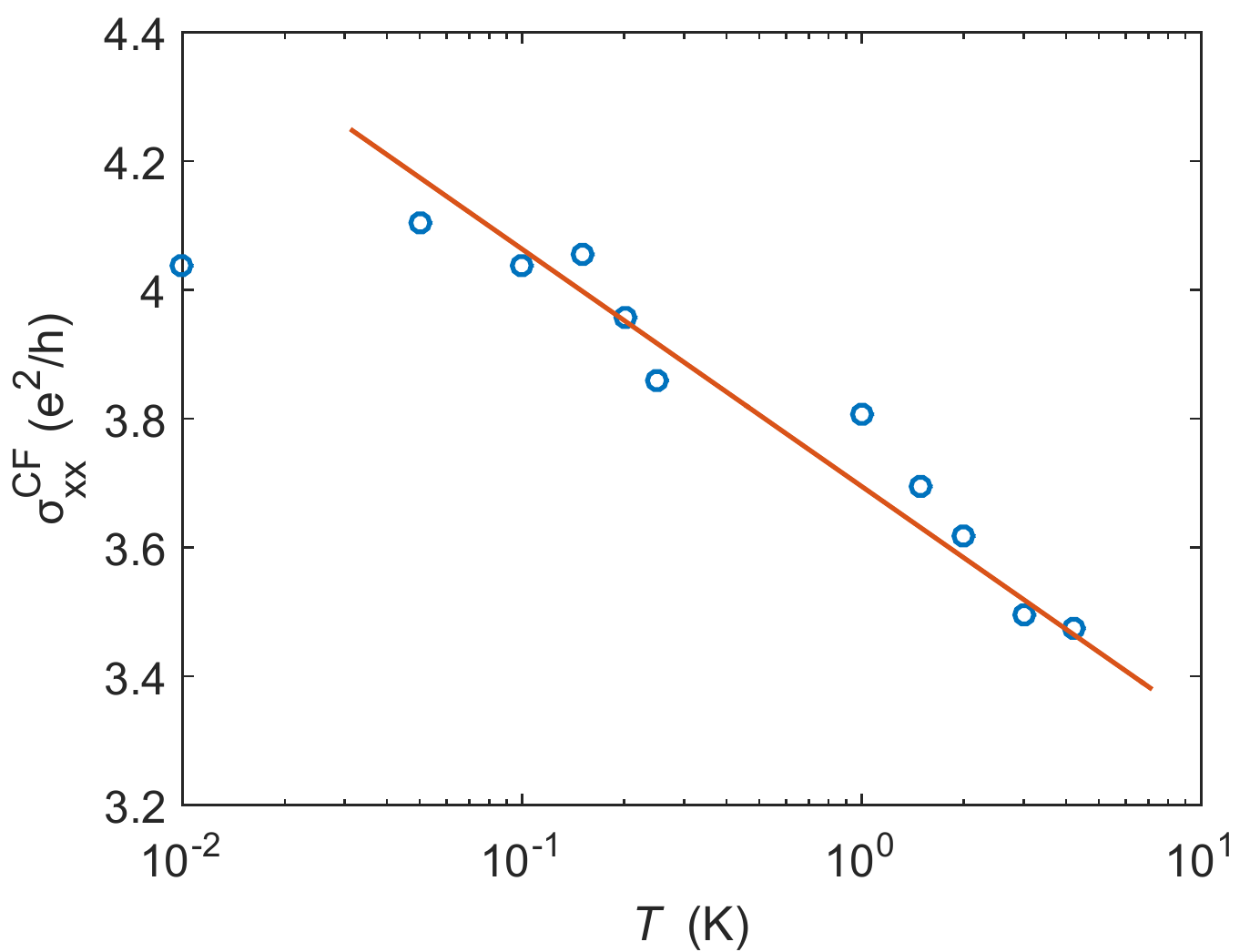}
\caption{a) Temperature dependence of conductivity $\sigma_{xx}$ for sample EV3  at half filling $\nu=1/2$. The fitted line corresponds to  logarithmic quantum correction with $\lambda =0.005$. b)  Temperature dependence of composite fermion conductivity $\sigma_{xx}^{CF}$ vs. $T$ at $\nu=1/2$; here the logarithmic correction amounts to  $\lambda = -0.16$ (see text).}
\label{G2d}
\end{figure}

Fig. \ref{G2d}a displays temperature dependence of conductivity $\sigma_{xx}$ for sample EV2 at $\nu=1/2$. The data indicate $\Delta \sigma_{xx} = \lambda \frac{e^2}{h}\log T/T_0$ \textit{increase} in $\sigma_{xx}$ with temperature where $\lambda = 0.005$; here $\Delta \sigma_{xx}=\sigma_{xx}(T)-\sigma_{xx}(T_0)$ with the reference temperature taken as $T_0=0.1$ K.  Altogether, the observed quantum correction is quite large and it amounts to $\sim20$\% of $\sigma_{xx}$ over the measured range.

We assume that particle-hole symmetry is valid in our 2D graphene electron gas experiment \cite{Kivelson1997,Nagaosa1998,Son2015}. Then, the connection between electronic conductivity $\sigma_{xx}$ in units $e^2/h$ and the composite fermion conductivity $\sigma^{CF}_{xx}$ is given by
\begin{equation}\label{sigmaconv}
  \sigma_{xx}= \frac{1}{4} \frac{\sigma^{CF}_{xx}}{{\sigma^{CF}_{xx}}^2 + {\sigma^{CF}_{xy}}^2} .
\end{equation}
Eq. \ref{sigmaconv} has been employed  to extract the diagonal conductivity of composite fermions $\sigma _{xx}^{CF}$.

The inversion of Eq. \ref{sigmaconv}  yields two solutions for $\sigma _{xx}^{CF}$. We retain only the metallic solution which reads as $\sigma _{xx}^{CF} = 1/4 \sigma _{xx}-\sigma _{xx}$, where we have used the fact $\sigma_{xy}^2=1/4$ for the particle-hole symmetric model and that $(4\sigma_{xx})^2 << 1$ for our data. The resulting conductivity $\sigma _{xx}^{CF}$ for composite fermions is displayed in Fig. \ref{G2d}b. Again, logarithmic dependence is obtained but now with $\lambda = -0.16$, which has a different sign when compared to previous observations on 2D electron gas systems with composite fermions at $\nu=1/2$ \cite{Kang1995,Rokhinson1995,Rokhinson1997}. One possible explanation for this different behavior is that Chern-Simon field fluctuations suppress very strongly weak localization contributions in regular 2D electron gas, leaving only interaction corrections to conductivity, while weak antilocalization effects still survive in graphene. In fact, the weak antilocalization properties of Dirac particles in graphene have  their particular characteristics due to the specific valley structure of graphene \cite{Mccann2006}.

In the presence of impurity scattering, the composite fermion motion will be
cut off by a finite transport mean free path $\ell$. For $q << 2/\ell$,
the random phase approximation result is ${\sigma _{xx}} = \frac{2}{{{k_F}\ell}}\frac{{{e^2}}}{{{{(2p)}^2}h}}$
where $2p$ indicates the number of vortices connected to each electron \cite{Jain2007}.
By fitting this formula to our data using $p=1$, we obtain $k_F \ell  \simeq 6.5$, which yields $\ell  \sim 150$ nm for the mean free path of composite fermions at $B = 5$ T. This indicates clearly diffusive motion as $\ell << L$. However, the impurity concentration in our sample does not coincide with this mean free path, and we conclude that this scattering length is set by the Chern-Simon field fluctuations.

The combined effect of the diffusive motion of composite fermions and the gauge field interaction has been studied by Khveshchenko \cite{Khveshchenko1996,Khveshchenko1997} and by Mirlin and W\"olfle \cite{Mirlin1997}. According to Khveshchenko \cite{Khveshchenko1997}, interference between disorder scattering and gauge fields leads to magnetoconductivity  $\sigma _{xx}^{CF}(B^*)$ with effective $B^*$:

\begin{equation}\label{magncond}
{\sigma _{xx}^{CF}(B^*) - \sigma _{xx}^{CF}(0)} =  - \delta {({\Omega _C}{\tau _{tr}})^2}\sigma _{xx}^{CF}(0)
\end{equation}
where $\Omega_C$ is the cyclotron frequency at effective field $B^* $, $\tau _{tr}$ denotes the transport scattering time and $\delta  = \frac{{\Delta \sigma _{xx}^{CF}(0)}}{{\sigma _{xx}^{CF}(0)}} $ is related to the sign of quantum corrections to conductivity, i.e.   $\delta  > 0$ and  $\delta  < 0$ for WAL and WL, respectively. Hence, weak antilocalization ($\lambda < 0$ in Fig. \ref{G2d}b) is tied with negative magnetoconductivity, which is in contrast to our observed positive sign.
One possible explanation for our positive magnetoconductivity in comparison with this model can be that Chern-Simon fluctuations display field dependence and they are reduced when going off from half filling, resulting in an increased mean free path $\ell$.
Consequently, the magnetoconductance becomes positive around $\nu=1/2$ but the temperature dependence at constant mean free path retains the antilocalization behavior.

\section{Discussion}
 Charge puddles and impurities create a disorder potential in suspended graphene, which can support localized FQH states \cite{Kumar2016}. The localized states can carry current across the sample  via coupling to adjacent localized states across the gapped FQH regions. However, since the energy gaps around $\nu=1/2$ appear smaller than their width $\simeq \Gamma_0$, there may be $\nu=1/2$ percolating paths around half filling, which guarantee that the nature of charge transport will not change sharply around $\nu=1/2$. There is  the possibility though that these percolation paths vary with magnetic field, which leads to sample dependent magnetoconductivity \cite{Mirlin1998}. We have cooled our samples a few times and found that the variation of magnetoconductivity at half filling remains  approximately within $\pm20\%$, but every time the sign of the quantum corrections have been the same at fields $7.5-9$T.

Our suspended graphene samples contain always built-in non-uniform strain, which is seen as frozen ripples at room temperature. This strain is modified by applied gate voltage, which can induce additional rippling around the perimeter if the graphene sheet is able to slide against the metallic contact. Variation in strain will lead to locally varying pseudomagnetic fields that will lead to TRS-breaking scattering within one cone (i.e. shorten $\tau_{*}$). According to Ref. \onlinecite{Couto2014}, non-uniform strain is the main contributor to both elastic scattering time $\tau$ and $\tau_{*}$, and $\tau \simeq \tau_{*}$ within a factor of 2-3. This scenario explains the  observation of WAL also in our sample at low magnetic fields.

Using a similar line of arguments as above, Chern-Simon fluctuations governing $\sigma_{xx}^{CF}$ will dominate both scattering rates $\tau^{-1}$ and $\tau_{*}^{-1}$ at $\nu=1/2$ at high fields.
Typically, it is argued that weak localization effects cannot be observed in the presence of Chern-Simon field fluctuations. This conclusion derives from the infrared divergence in the density of states of low frequency excitations at small wave vectors $q$ \cite{Simon1998}. In our sample, however, there is a rather high infrared cutoff due to small geometrical size and low carrier density. By taking $q_{min} \simeq 1/(r_o - r_i)$ and $k_F = \sqrt{\pi n}$, we obtain for the cut-off $q_{min}/k_F \simeq 0.01$. Consequently, the Chern-Simon fields do not kill weak localization effects of composite fermions in our graphene sample, but rather just lower the dephasing time (and the ratios $\tau_{\varphi}/\tau_*$  and $\tau_{\varphi}/\tau_i$) which favors weak antilocalization behavior among graphene Dirac particles \cite{Tikhonenko2009}.

\section{Conclusions}
We have investigated fractional quantum Hall states in suspended graphene Corbino disk around half filling.  We find  weak logarithmic temperature dependence of conductivity,  the sign of  which  indicates weak antilocalization behavior of graphene composite fermions. These observations with nearly zero effective field  acting on composite fermions can be understood making a parallel with graphene Dirac particles at small magnetic fields and assuming an enhanced infra-red cut-off for Chern-Simon fluctuations in a graphene sample of small carrier density and geometrical size.

\section{Acknowledgements}
We thank C. Flindt, A. Harju, Y. Meir, T. Ojanen, S. Paraoanu, and E. Sonin for fruitful discussions. This work has been supported in part by the EU Framework Programme (H2020 Graphene Flagship) and the European Research Council (grant no. 670743), and by the Academy of Finland (projects no. 250280 LTQ CoE and 286098).


%

\end{document}